\documentclass[multphys,vecphys]{svmult}

\usepackage{makeidx}         
\usepackage{graphicx}        
\usepackage{multicol}        
\usepackage[bottom]{footmisc}

\makeindex             
\newcommand       \beq          {\begin{equation}}
\newcommand       \eeq          {\end{equation}}

\newcommand       \cm           {\,{\rm cm}}

\newcommand       \g            {\,{\rm g}}

\newcommand       \K            {\,{\rm K}}

\newcommand       \km           {\,{\rm km}}

\newcommand       \s            {\,{\rm s}}

\newcommand       \yrs           {\,{\rm yrs}}

\newcommand       \simali       {\sim\,}
\newcommand       \hab          {{\rm HAeBe}}
\newcommand       \Msun         {\,{\rm M_{\odot}}}
\newcommand       \Teff         {T_{\rm eff}}
\newcommand       \Stardust     {{\it Stardust}}         

\newcommand       \mum          {\,{\rm \mu m}}
\newcommand	  \ppm		{\,{\rm ppm}}
\newcommand       \nm           {\,{\rm nm}}
\newcommand       \AU           {\,{\rm AU}}

\newcommand       \ch           {\rm C_{14}H_{10}}
\newcommand       \rh           {r_{\rm h}}

\begin{document}
\title*{PAHs in Comets: An Overview}
\author{Aigen Li}
\institute{Department of Physics and Astronomy,
           University of Missouri,\\ 
           Columbia, MO 65211;
\texttt{LiA@missouri.edu}}
%
%
\maketitle

%
%

\setcounter{footnote}{0}

%
%

\vspace{-0.4cm}
\begin{abstract}
\vspace{-0.8cm}
Polycyclic aromatic hydrocarbon (PAH) molecules,
ubiquitously seen in the interstellar medium (ISM) 
of our own and external galaxies,  
might have been incorporated into comets if they are 
formed from relatively unprocessed interstellar matter.
The detection of PAHs in comets would be 
an important link between the ISM and comets.
This review compiles our current knowledge on
cometary PAHs, based on ground-based and space-borne
observations of infrared vibrational and ultraviolet
fluorescence spectra of comets, and laboratory analysis of
interplanetary dust particles possibly of cometary origin 
and cometary samples returned to Earth by the \Stardust\ spacecraft. 
The latter provided the most unambiguous evidence
for the presence of PAHs in cometary nuclei.    
\end{abstract}

\vspace{-1.2cm}
\section{~Introduction: PAHs as a Link between 
         the Interstellar Medium and the Solar System}
\vspace{-0.2cm}
The major goals of cometary science are to determine 
the chemical composition and physical structure of 
cometary nuclei and to shed light on the origin of
the solar system. It is now widely recognized that
comets formed in the cold outer regions of 
the solar nebula ($\simali$5--55$\AU$ from the Sun, 
well beyond the ``snowline'') and have been stored 
in two distant reservoirs (i.e. the Oort cloud and 
the Kuiper Belt) for most of the age of the solar system
($\simali$4.6$\times 10^9\yrs$).
Because of their cold formation and cold storage,\footnote{%
  Long-period comets (with orbital periods ranging from 
  200$\yrs$ up to $10^7\yrs$) originate in the Oort Cloud 
  ($\simali$3000--50,000$\AU$ from the Sun)
  and formed in the giant planets region 
  from Jupiter to Neptune
  of the pre-planetary nebula ($\simali$5--40$\AU$ from the Sun), 
  where the nebula temperature ranged from $>$120\,K near Jupiter 
  to $<$30\,K near Neptune.
  The source of Jupiter-family short-period comets 
  (with periods shorter than 20$\yrs$),
  formed further out than long-period comets 
  in the trans-Neptune region,
  is the Kuiper Belt
  ($\simali$30--50$\AU$ from the Sun).
  Halley-family short-period comets 
  (with periods of 20$\yrs$\,$<$\,$P$\,$<$\,200$\yrs$)
   originally came from the Oort cloud 
   and then have been scattered into short-period type orbits 
   by the perturbation of Jupiter and/or Saturn.
   }
it is therefore also widely believed that comets 
are the most primitive objects in the solar system.

However, there is no consensus on to what 
extent comets preserve the composition of 
the presolar molecular cloud 
and the early stages of the protosolar nebula. 
A compelling theory is that comets are made of 
unaltered pristine interstellar materials 
with only the most volatile components partially 
evaporated (Greenberg 1982). 
Alternatively, it has also been proposed that 
cometary materials have been subjected to evaporation, 
recondensation and other reprocessing in the protosolar 
nebula and therefore have lost all the records of 
the presolar molecular cloud out of which they have formed.

\begin{figure}
\centering
\includegraphics[width=11.6cm]{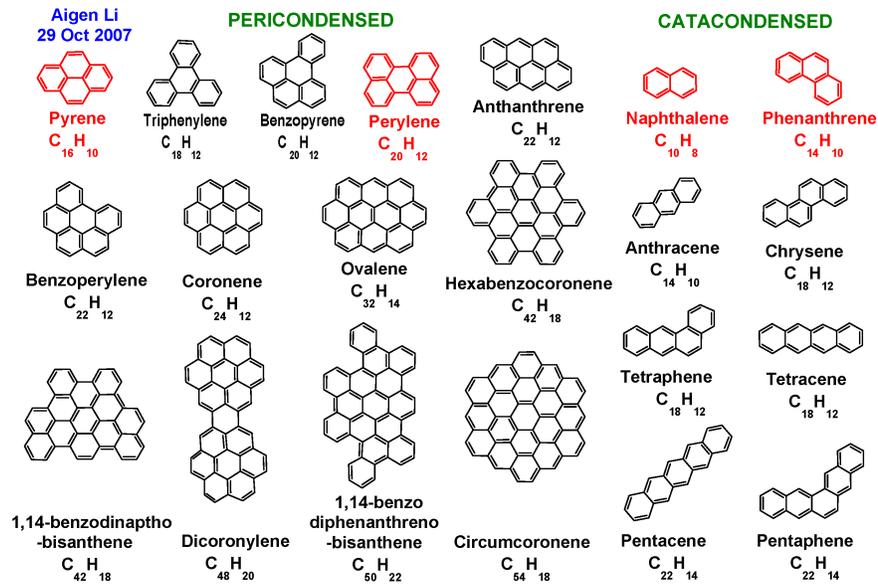}
\vspace{-0.2cm}
\caption{\label{fig:pah_cartoon}\footnotesize 
         Structures of 21 {\it specific} PAH molecules.
         Both compact pericondensed molecules
         ({\it pyrene}, {\it perylene}) 
         and thermodynamically less favoured catacondensed 
         molecules ({\it naphthalene}, {\it phenanthrene})
         and their alkylated homologs
         were identified in the \Stardust\ samples
         (Sandford et al.\ 2006, Clemett et al.\ 2007).  
         Naphthalene, phenanthrene and their alkylated derivatives
         were found in IDPs possibly of cometary origin
         (Clemett et al.\ 1993). 
         PAHs with up to 7 rings and their alkyl derivatives
         are abundant in carbonaceous chondrites
         (Sephton et al.\ 2004).
         }
\end{figure}

Polycyclic aromatic hydrocarbon (PAH) molecules,
composed of fused benzene rings 
(see Fig.\,\ref{fig:pah_cartoon} for illustration),
and a significant constituent of 
the interstellar medium (ISM) of the Milky Way
and external galaxies (see \S\ref{sec:ism}), 
would also be present in comets if they 
indeed contain unprocessed interstellar matter.
The detection of PAHs in comets would be an important
link between the ISM and comets and provides important 
clues on the processes that occurred during the formation 
of our solar system. 

In this review I attempt to compile all possible 
evidence for the presence of PAHs in comets (\S\ref{sec:cometpah}), 
focusing on ground-based and space-borne spectroscopy
of infrared vibrational and ultraviolet (UV)
fluorescent emission spectra of comets, 
and laboratory analysis of stratospherically collected
interplanetary dust particles (IDPs) 
thought to be cometary in origin 
and cometary samples returned to Earth 
by the \Stardust\ spacecraft,
with the latter providing the most unambiguous proof.
Using the PAHs in the ISM (\S\ref{sec:ism}) 
and circumstellar disks (\S\ref{sec:disk}) 
as a comparison basis,
the physical and chemical nature 
and source of cometary PAHs are discussed in \S\ref{sec:discussion}.

\vspace{-0.1cm}

\section{~Ubiquity of PAHs in the Interstellar Medium of 
          the Milky Way and External Galaxies\label{sec:ism}}
\begin{figure}
\centering
\includegraphics[height=11cm]{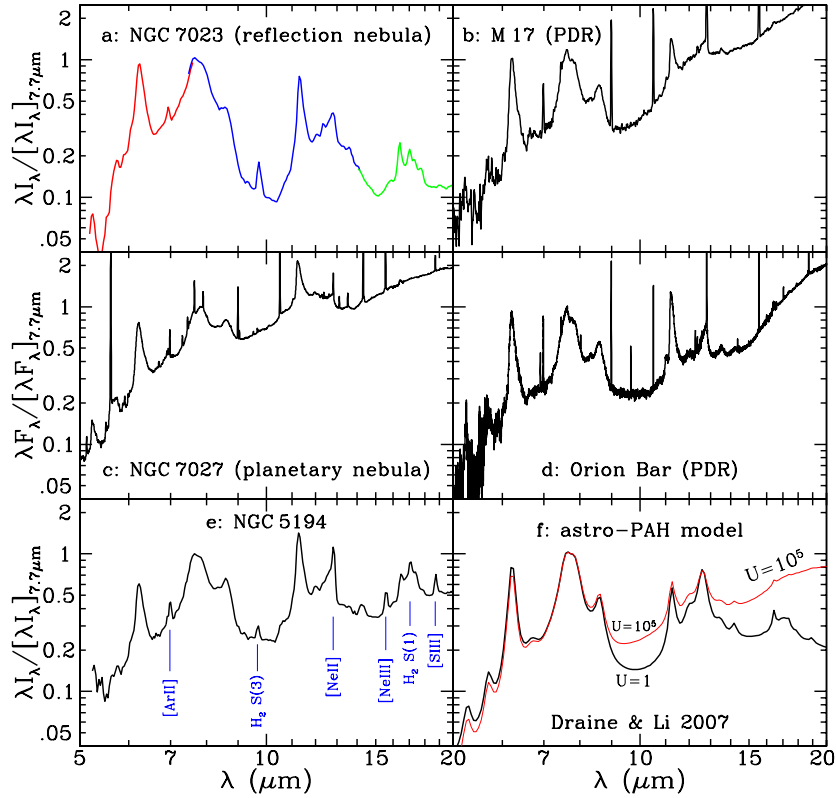}
\vspace{-3mm}
\caption{\label{fig:pah_ism} \footnotesize
         Observed 5--20$\mum$ spectra for: 
	 (a) Reflection nebula NGC\,7023
	     (Werner et al.\ 2004);
	 (b) Orion Bar photodissociated region
             (PDR; Verstraete et al.\ 2001);
	 (c) M17 PDR
             (Peeters et al.\ 2005);
	 (d) Planetary nebula NGC\,7027
	     (van Diedenhoven et al.\ 2004);
	 (e) Seyfert Galaxy NGC\,5194
             (Smith et al.\ 2007).
	 Also shown (f) is the emission calculated for
	 the astro-PAH model of Draine \& Li (2007),
         illuminated by starlight with an intensity
         $U=1$ and $10^5$ times of the local interstellar
         radiation field of Mathis et al.\ (1983). 
         Taken from Draine \& Li (2007).
	 }
\end{figure}

\vspace{-0.2cm}
PAHs reveal their presence in the ISM
by emitting a distinctive set of emission features 
at 3.3, 6.2, 7.7, 8.6, and 11.3$\mum$ 
(which are also collectively known as 
the ``Unidentified Infrared'' [UIR] emission bands).\footnote{%
  \vspace{-0.5mm} 
  These ``UIR'' emission features 
  are now generally identified as vibrational modes of 
  PAHs (L\'{e}ger \& Puget 1984; Allamandola et al.\ 1985):
  C--H stretching mode (3.3$\mum$), 
  C--C stretching modes (6.2, 7.7$\mum$), 
  C--H in-plane bending mode (8.6$\mum$),
  and C--H out-of-plane bending mode (11.3$\mum$).
  Other C--H out-of-plane bending modes 
  at 11.9, 12.7 and 13.6$\mum$ have also been detected.
  The wavelengths of the C--H out-of-plane bending modes 
  depend on the number of neighboring H atoms:
  11.3$\mum$ for solo-CH (no adjacent H atom),
  11.9$\mum$ for duet-CH (2 adjacent H atoms),
  12.7$\mum$ for trio-CH (3 adjacent H atoms),
  and 13.6$\mum$ for quartet-CH (4 adjacent H atoms).
  Other prominent features are the C-C-C bending modes
  at 16.4$\mum$ (Moutou et al.\ 1996),
  17.1, 17.8, 18.9$\mum$ 
  (Beintema et al.\ 1996; 
   Smith et al.\ 2004; 
   van Kerckhoven et al.\ 2000; 
   Werner et al.\ 2004).
   }
Since their first detection in the planetary nebulae NGC\,7027 
and BD+30$^{\rm o}$3639 (Gillett et al.\ 1973), 
PAHs have been observed in a wide range of
Galactic and extragalactic regions
(see Fig.\,\ref{fig:pah_ism} and Draine \& Li 2007). 

In the Milky Way diffuse ISM, 
PAHs, containing $\simali$45$\ppm$ 
(parts per million, relative to H) C,
account for $\simali$20\% of the total power emitted 
by interstellar dust (Li \& Draine 2001b).
The {\it\small ISO} (Infrared Space Observatories) 
and {\it Spitzer} imaging and spectroscopy 
have revealed that PAHs are also a ubiquitous feature of 
external galaxies (Tielens 2005; Smith et al.\ 2007). 
Recent discoveries include 
the detection of PAH emission in a wide range of systems:
distant Luminous Infrared Galaxies (LIRGs) 
with redshift $z$ ranging from 0.1 to 1.2 
(Elbaz et al.\ 2005),
distant Ultraluminous Infrared Galaxies 
(ULIRGs) with redshift $z\sim$\,2 (Yan et al.\ 2005, 2007),
distant luminous submillimeter galaxies at 
redshift $z\sim$\,2.8 (Lutz et al.\ 2005),
the distant Cloverleaf lensed QSO 
at redshift $z\sim$\,2.56 (Lutz et al.\ 2007),
elliptical galaxies with a hostile environment
(containing hot gas of temperature $\sim$\,10$^7\K$) 
where PAHs can be easily destroyed 
through sputtering by plasma ions
(Kaneda et al.\ 2005),
faint tidal dwarf galaxies with metallicity 
$\sim Z_\odot/3$ (Higdon et al.\ 2006),
and galaxy halos 
(extending $\simali$6.5\,kpc 
from the plane of NGC\,5907 [Irwin \& Madden 2006]
and $>$9.5\,kpc from the plane of M\,82
[Engelbracht et al.\ 2006]). 

However, the PAH features are weak or even absent 
in AGNs (as first noticed by Roche et al.\ 1991)
and low-metallicity galaxies
(Thuan et al.\ 1999; Houck et al.\ 2004;
Engelbracht et al.\ 2005; Hunt et al.\ 2005;
Madden et al.\ 2006; Wu et al.\ 2006;
Rosenberg et al.\ 2006; Draine et al.\ 2007).
The exact reason for the deficiency or lack of PAHs 
in low-metallicity galaxies and AGNs is not clear. 
It is generally interpreted as the destruction of PAHs 
(1) by hard UV photons (e.g. see Plante \& Sauvage 2002)
    or supernova-driven shocks (O'Halloran et al.\ 2006) 
    in metal-poor galaxies,\footnote{%
     Alternatively, these low-metallicity galaxies 
     may be truly young and PAHs have simply not had 
     time to form due to the delayed injection of carbon 
     molecules from low-mass stars into the ISM (Dwek 2005).
     }
and 
(2) by extreme UV and soft X-ray photons in AGNs 
    (Voit 1991, 1992; Siebenmorgen et al.\ 2004).
Observations also suggest that PAHs are destroyed in 
star-forming regions with very strong and hard radiation fields 
(Contursi et al. 2000; F\"{o}rster Schreiber et al.\ 2004; 
Beir\~{a}o et al.\ 2006).

\vspace{-2mm}

\vspace{-12mm}
\section{~PAHs in Circumstellar Dust Disks\label{sec:disk}} 
\begin{figure}
\vspace{-3mm}
\centering
\includegraphics[height=11cm]{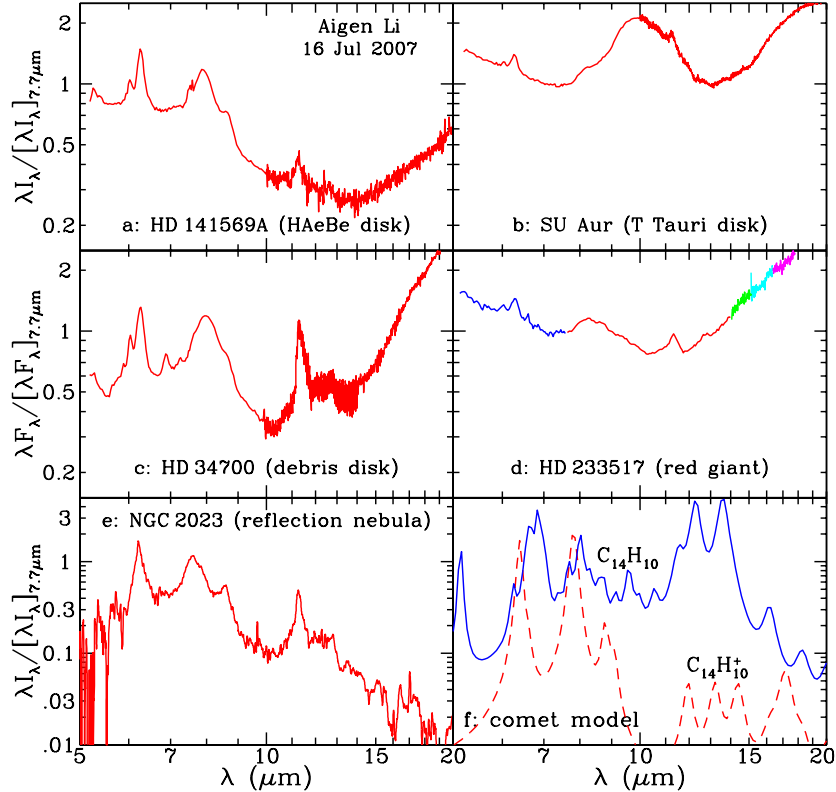}
\vspace{-5mm}
\caption{\label{fig:pah_disks} \footnotesize
         Observed 5--20$\mum$ spectra for: 
	 (a) Protoplanetary disk around HAeBe star 
             HD\,141569A (B9.5V; $\Teff$\,$\approx$\,10,000$\K$;
             Sloan et al.\ 2005);
	 (b) Protoplanetary disk around T Tauri star 
             SU Aur (G1III; $\Teff$\,$\approx$\,5945$\K$;
             Furlan et al.\ 2006);
	 (c) Debris disk around HD\,34700
             (G0V; $\Teff$\,$\approx$\,6000$\K$; 
              Li et al.\ 2008);
	 (d) Circumstellar disk around red giant
             HD\,233517 (K2III; $\Teff$\,$\approx$\,4390$\K$; 
             Jura et al.\ 2006);
	 (e) Reflection nebula NGC\,2023
             (illuminated by HD\,37903 
              [B1.5V; $\Teff$\,$\approx$\,22,000$\K$];
               Verstraete et al.\ 2001).
	 Also shown (f) is the emission calculated for
	 phenanthrene C$_{14}$H$_{10}$
         and its cation C$_{14}$H$_{10}^{+}$ at 
         $\rh$\,=\,1$\AU$ from the Sun (Li \& Draine 2008).
	 }
\vspace{-2mm}
\end{figure}

\vspace{-1mm}
Dust disks around young stars, depending on their age,
are the source material or the remnants of 
newly-formed planets, asteroids, and comets.
There exists observational evidence for 
the presence of PAHs in protoplanetary disks 
around the intermediate-mass ($\simali$2--10$\Msun$)
pre-main-sequence (PMS) 
Herbig Ae/Be (hereafter HAeBe) stars 
and their low-mass ($<$\,2$\Msun$) analog T Tauri stars, 
as well as debris disks around main-sequence (MS) stars
(see Fig.\,\ref{fig:pah_disks} for illustration).

\begin{itemize}
\vspace{-0.3em}
\item From an analysis of the space-borne and ground-based
      spectra of 41 \hab\ stars in the 3$\mum$ region,
      Brooke et al.\ (1993) reported a firm detection 
      of the 3.3$\mum$ PAH C--H stretching emission feature 
      in $\simali$20\% of these objects. 
\vspace{-0.5em}
\item Acke \& van den Ancker (2004) found that the PAH features 
      have been detected in $\simali$57\% of the 46 \hab\ stars 
      for which the {\it\small ISO} spectroscopic data are available.
\vspace{0.4em}
\item Recent {\it Spitzer} observations have 
      obtained the PAH spectra of over 20 \hab\ stars 
      (Sloan et al.\ 2005; Keller et al.\ 2008). 
      Spectral variations among these stars 
      and their deviations from typical interstellar 
      PAH features were reported
      (see Fig.\,\ref{fig:pah_disks}).
\vspace{0.4em}
\item Geers et al.\ (2006) analyzed the {\it Spitzer} IRS spectra 
      of 38 T Tauri stars and found PAHs in at least 8\% 
      (or probably as much as 45\%) of these objects.
       The PAH spectra of T Tauri stars appear to be quite 
       different from those of \hab\ stars 
       and those typical in the ISM
       (see Fig.\,\ref{fig:pah_disks}).
%
%
\vspace{0.4em}
\item The PAH emission features have also been detected in
      UV-poor dust debris disks around F- and G-type MS stars 
      (e.g. SAO\,206462 of spectral type F8V with
       $\Teff\approx 6250\K$ [Coulson \& Walther 1995],
       and HD\,34700 of spectral type G0V
       with $\Teff\approx 5940\K$ 
       [Sylvester et al.\ 1997; 
        Smith, Clayton, \& Valencic 2004;
        see Fig.\,\ref{fig:pah_disks}]).
      However, in an extensive {\it Spitzer} IRS spectroscopic
      survey of 111 T Tauri stars in the Taurus star-forming region,
      Furlan et al.\ (2006) found that the PAH emission
      bands are not seen in dust disks around T Tauri stars of 
      spectral type later than G1. 
\vspace{0.4em}
\item Ground-based spatially resolved spectroscopy 
      has revealed that the 3.3--12.7$\mum$ PAH emission 
      features in some HAeBe disks are spatially extended 
      (on a scale of several hundred AU for the 6.2--12.7$\mum$ 
       bands; van Boekel et al.\ 2004; Geers et al.\ 2004; 
       Habart et al.\ 2006). 
\vspace{0.4em}
\item Jura et al.\ (2006) reported the detection
      of the PAH emission features in HD\,233517, 
      an evolved oxygen-rich K2III red giant 
      ($\Teff \approx 4390\K$) with circumstellar dust. 
      But Jura (2003) argued that 
      the IR excess around HD\,233517 is unlikely to 
      be produced by a recent outflow in a stellar wind. 
      Jura et al.\ (2006) hypothesized that
      there is a passive, flared disk orbiting HD\,233517
      and the PAH molecules in the orbiting disk 
      may be synthesized in situ as well as having been 
      incorporated from the ISM.\\[1mm]
      Sloan et al.\ (2007) observed the PAH emission features
      in the circumstellar disk of HD\,100764,
      a carbon-rich red giant ($\Teff \approx 4850\K$),
      and found that they are shifted to longer wavelengths
      than normally seen, consistent with a ``Class C'' 
      PAH spectrum (Peeters et al.\ 2002; 
      the PAH spectra of HD\,233517 and SU Aur also 
      belong to ``Class C''; see Fig.\,\ref{fig:pah_disks}).
%
%
\end{itemize}

\begin{figure}
\vspace{-3mm}
\centering
\includegraphics[height=7.1cm]{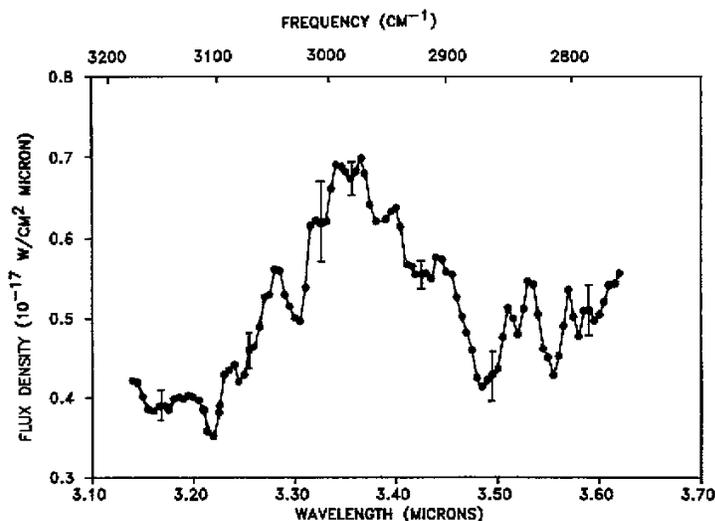}
\vspace{-4mm}
\caption{\label{fig:halley_3.28um} \footnotesize
         IR emission spectrum of comet Halley 
         obtained by Baas et al.\ (1986) with 
         the 3.8\,m United Kingdom IR Telescope (UKIRT)
         on Mauna Kea on 1986 April 25 
         at a heliocentric distance of $\rh$\,=\,1.6$\AU$.
         In addition to the major feature at 3.36$\mum$
         (often known as the ``cometary organic feature'';
          the 3.33$\mum$ $\nu_2$ band and 3.37$\mum$ $\nu_9$ band 
          of methanol CH$_3$OH account for about half its total 
          intensity; Bockel\'ee-Morvan et al.\ 1995),
         there were also subsidiary peaks at 3.28$\mum$
         and 3.52$\mum$ ($\nu_3$ band of methanol; Hoban et al.\ 1993). 
         Taken from Baas et al.\ (1986).
	 }
\end{figure}

\begin{figure}
\vspace{-2mm}
\centering
\includegraphics[height=7.1cm]{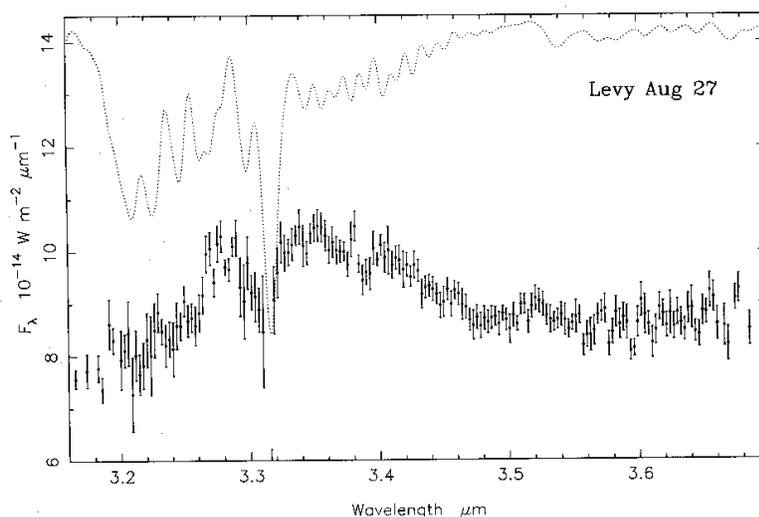}
%
\vspace{-4mm}
\caption{\label{fig:levy_3.28um} \footnotesize
         IR emission spectrum of comet Levy
         obtained by Davies et al.\ (1991) with UKIRT
         on 1990 August 27 at $\rh$\,=\,1.4$\AU$.
         The 3.28$\mum$ feature was as prominent
         as the broad 3.4$\mum$ (which was resolved 
         into two components peaking at 3.35$\mum$
         and 3.41$\mum$). The dotted curve was
         the relative transmission of 
         the atmosphere above Mauna Kea.
         Comet Levy had the strongest 3.28$\mum$ feature
         relative to the 3.4$\mum$ feature among all comets.
         Taken from Davies et al.\ (1991).
	 }
\end{figure}

\begin{figure}
\vspace{-3.5mm}
\centering
\includegraphics[height=7.0cm]{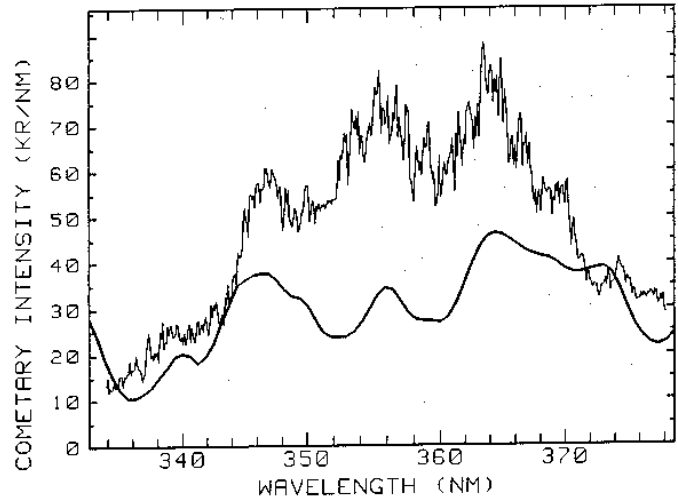}
\vspace{-3.5mm}
\caption{\label{fig:moreels} \footnotesize
         Comparison of the near-UV emission spectrum
         of comet Halley measured by TKS-Vega at 
         $\rh$\,=\,0.83$\AU$ (on 1986 March 9)
         with the experimentally-measured 
         laser-induced fluorescence spectrum of
         phenanthrene (upper curve).
         Three main peaks coincide at 347, 356, and 364\,nm.
         Taken from Moreels et al.\ (1994).
	 }
\vspace{-5mm}
\end{figure}

\begin{figure}
\vspace{-3.2mm}
\centering
\includegraphics[width=11.2cm]{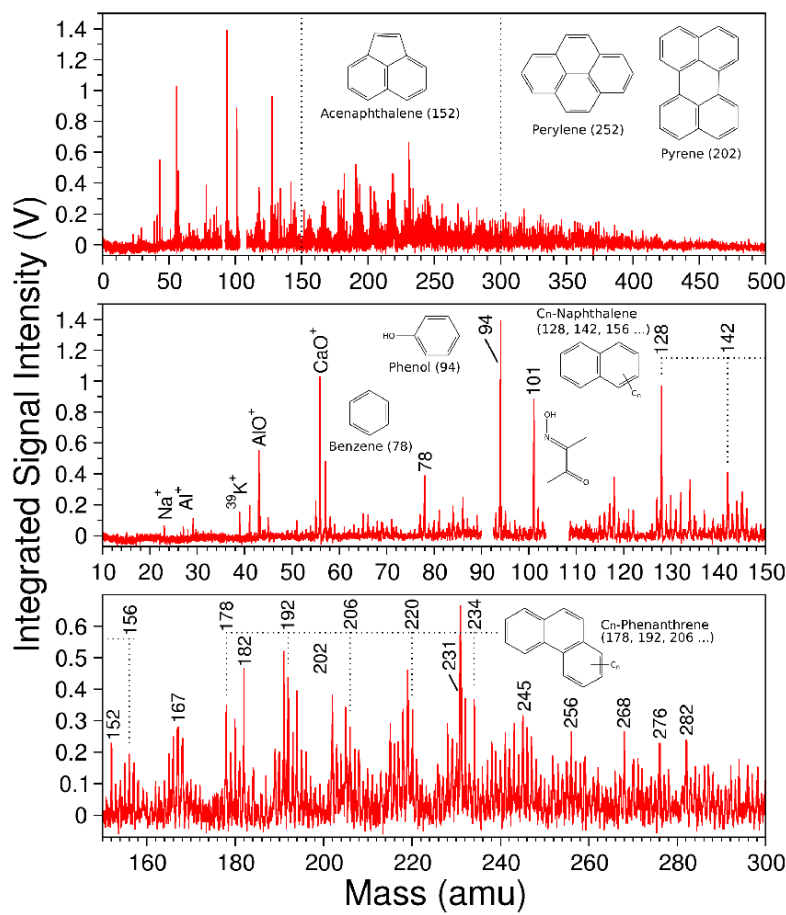}
\vspace{-2mm}
\caption{\label{fig:clemett} \footnotesize
         PAH mass spectrum distribution for a \Stardust\ sample
         obtained with the two-step laser mass spectrometry.
         The most commonly found PAH species are 
         naphthalene (C$_{10}$H$_8$),
         phenanthrene (C$_{14}$H$_{10}$),
         pyrene (C$_{16}$H$_{10}$),
         perylene (C$_{20}$H$_{12}$),
         and their alkylated homologs.
         Interspersed within these species is
         a rich suite of auxiliary peaks which appears to
         represent the presence of O and N substitution, 
         where the heterofunctionality being
         external to aromatic structure.
         Taken from Clemett et al.\ (2007).
	 }
\vspace{-6mm}
\end{figure}

\vspace{-7mm}
\section{~Evidence for PAHs in Comets\label{sec:cometpah}} 
\vspace{-3mm}
The presence of PAHs in comets was an open issue
in the pre-\Stardust\ era. One of the most important 
discoveries of the \Stardust\ Discovery Mission was
the first ever most unambiguous detection of PAHs in 
a comet. Below, I summarize all {\it tentative} evidence
in the pre-\Stardust\ era which may suggest the presence
of PAHs in comets and finally, the more definite proof
from the \Stardust\ mission.\footnote{%
   We should note that PAHs (e.g. naphthalene, 
   phenanthrene, pyrene, chrysene, perylene, 
   benzo[ghi]perylene, and coronene; 
   see Fig.\,\ref{fig:pah_cartoon}) with varying degrees 
   of alkylation have been identified in primitive 
   carbonaceous chondritic meteorites
   (Sephton et al.\ 2004; Derenne et al.\ 2005).
   Although most primitive meteorites are {\it asteroidal}
   (originated in the asteroid belt, somewhere between
    2--5$\AU$ from the Sun), Campins \& Swindle (1998)
   argued that some meteorites may have a {\it cometary} origin
   (but so far no known meteorites appear to come from comets).
   }
\begin{itemize}
\item Baas et al.\ (1986) by the first time reported 
      the detection of a discrete emission feature
      at 3.28$\mum$ in comet Halley after perihelion,
      at heliocentric distances of $\rh$\,=\,1.6$\AU$
      (on 1986 April 25; see Fig.\,\ref{fig:halley_3.28um}) 
      and 2.0$\AU$ (on 1986 May 24).
      This feature was also seen in comet Levy 
      at $\rh$\,=\,1.4$\AU$ 
      (on 1990 August 27; Davies et al.\ 1991; 
       see Fig.\,\ref{fig:levy_3.28um}). 
      It was tentatively attributed to the C--H stretching mode 
      of PAHs (e.g. see Bockelee-Morvan et al.\ 1995).
      However, other species, such as CH$_4$ and OH prompt emission 
      are also contributing at this wavelength
      (e.g. see Mumma et al.\ 2001).\\[1mm]
      %
%
      Using a $\chi^2$-fitting technique, 
      Lisse et al.\ (2006) found the 6.2, 7.7, 8.6 
      and 11.3$\mum$ emission features of {\small PAHs} 
      in the spectrum of the {\it Deep Impact} ejecta 
      of comet Tempel 1 (at $\rh$\,=\,1.51$\AU$),
      obtained with {\it Spitzer} 45 minutes after impact.
      This identification did not rely on a search for
      individual spectral features, but from the decrease
      of the residual $\chi^2$ after a fit of the observed
      spectrum by a modeled spectrum.\footnote{%
        In the ground-based 7.8--13.2$\mum$ mid-IR spectra 
        of comet Tempel 1 obtained with {\it Gemini-N} 
        61--94 minutes after impact,
        Harker et al.\ (2007) possibly detected emission 
        from PAHs at 8.25 and 8.6$\mum$ 
        (perhaps also at 12.4 and 12.9$\mum$).
        It is unclear whether the 11.3$\mum$ PAH feature
        was present or not since it might have been hidden by
        the much stronger 11.3$\mum$ crystalline olivine feature.
        }
      Using the same technique, Lisse et al.\ (2007) re-analyzed
      the {\small\it ISO} spectrum of Hale-Bopp at $\rh$\,=\,2.8$\AU$
      taken on 1996 October 6 and found PAH signals in this
      \vspace{-6mm} 
      comet (but see Crovisier et al.\ 2000, 
      Crovisier \& Bockel\'{e}e-Morvan 2008).
      %
%
\vspace{0.6em}
\item Moreels et al.\ (1994) reported the detection of
      four emission bands at 347, 356, 364 and 374$\nm$
      in the near-UV spectrum of comet Halley 
      at $\rh$\,=\,0.83$\AU$ obtained with 
      the Three-Channel-Spectrometer (TKS) 
      on board the Vega-2 spacecraft (see Fig.\,\ref{fig:moreels}).
      They attributed these bands to the fluorescence 
      of phenanthrene ($\ch$).
      But the production rate of phenanthrene alone 
      required to account for the observed band intensities  
      ($\simali$1.5\,$\times$\,10$^{27}$\,mol\,$\s^{-1}$)       
      would be $\simali$100 times higher than 
      that of PAHs derived from the 3.28$\mum$ feature
      (Bockel\'{e}e-Morvan et al.\ 1995). 
      More recently, Clairemidi et al.\ (2004) identified
      in the TKS-Vega spectrum of the inner coma of comet Halley
      a broad-band emission feature between 340 and 390$\nm$
      with 3 peaks at 371, 376 and 382$\nm$. 
      They tentatively attributed these bands to 
      pyrene (C$_{16}$H$_{10}$).
\vspace{0.3em}
\item Clemett et al.\ (1993) identified small PAH molecules
      (including naphthalene C$_{10}$H$_8$ and phenanthrene)
      and their alkylated derivatives in the microprobe 
      two-step laser desorption laser ionization mass 
      spectrometry ($\mu$L$^2$MS) of 
      IDPs possibly of cometary origin.\footnote{%
        Stratospheric IDPs are believed to have originated 
        primarily from asteroids and short-period comets
        (i.e. collisional debris from main belt asteroids 
         and cometary dust captured in Earth's atmosphere
         when spiraling toward the Sun due to 
         Poynting-Robertson drag).
        Atmospheric entry velocities,
        as determined from the atmospheric entry 
        temperatures measured from the stepped He-release 
        method (Nier \& Schlutter 1993) 
        and the atmospheric entry model of
        Love \& Brownlee (1994),
        have been used to distinguish between IDPs arising
        from comets and asteroids, based on
        the marked differences between typical
        asteroidal and cometary orbits
        -- asteroidal IDPs spiraling in toward the Sun 
        from low inclination, low eccentric asteroidal sources 
        will enter Earth's atmosphere, on average, 
        at relatively lower velocities
        than cometary IDPs (Flynn 1989, Joswiak et al.\ 2007).	
        Very recently, Joswiak  et al.\ (2007) analyzed 31 IDPs 
        and distinguished two groups:
        12 porous cometary IDPs with atmospheric entry velocities
        $>$\,18$\km\s^{-1}$, an average density of
        $\simali$1.0$\g\cm^{-3}$, and an anhydrous mineralogy;
        4 more compact asteroidal IDPs mainly composed 
        of hydrated minerals with atmospheric entry 
        velocities $<$\,14$\km\s^{-1}$, and an average density of
        $\simali$3.3$\g\cm^{-3}$.
        }
\vspace{0.3em}
\item Very recently, Sandford et al.\ (2006),
      Keller et al.\ (2006), and Clemett et al.\ (2007)
      analyzed the cometary materials
      returned to Earth by the \Stardust\ spacecraft
      and clearly demonstrated that PAHs are present
      in comet Wild 2, the \Stardust\ mission target:
      (1) The L$^2$MS (two-step laser desorption laser ionization 
          mass spectrometry) spectra of the \Stardust\ samples
          revealed the presence of 
          naphthalene, phenanthrene, pyrene, 
          perylene (C$_{20}$H$_{12}$)
          and their alkylated homologs extending up to
          at least C$_4$-alkyl 
          (Sandford et al.\ 2006, Clemett et al.\ 2007; 
           Fig.\,\ref{fig:clemett});
       (2) The Raman spectra of the \Stardust\ samples
           exhibit pronounced ``D'' ($\simali$1360\,${\rm \Delta}\cm^{-1}$)
           and ``G'' ($\simali$1580\,${\rm \Delta}\cm^{-1}$) bands,
           characteristic of highly disordered sp$^2$-bonded 
           aromatic carbon (Sandford et al.\ 2006; 
           Fig.\,\ref{fig:Raman_FTIR}A); and
       (3) The Fourier transform infrared (FTIR) spectra
           clearly show an absorption feature at 3050$\cm^{-1}$,
           corresponding to the aromatic C--H stretching mode
           at 3.28$\mum$ (Keller et al.\ 2006; 
           Fig.\,\ref{fig:Raman_FTIR}B). 
          %
%
\end{itemize}



\vspace{-4.5mm}
\begin{figure}
\vspace{-4.5mm}
\centering
\includegraphics[width=11.5cm]{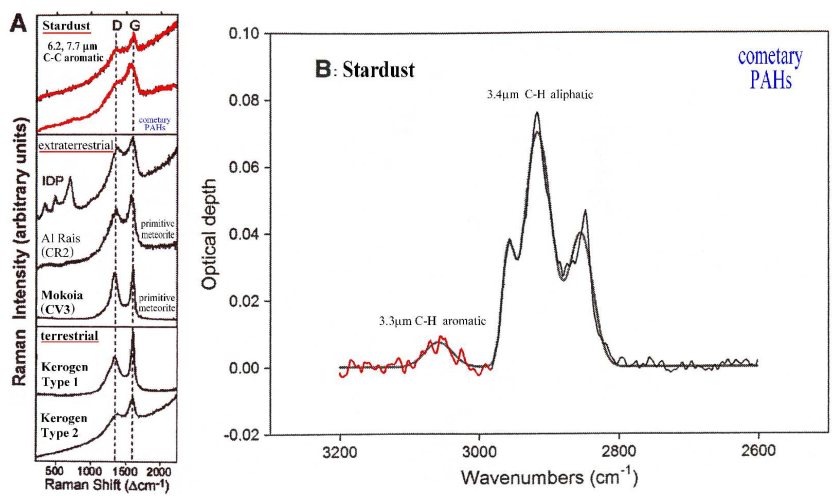}
\vspace{-0.5mm}
\caption{\label{fig:Raman_FTIR} \footnotesize
         {\bf Left (A)}: Raman spectra of two \Stardust\ particles 
         (top panel) compared with the spectra of organics from 
         extraterrestrial 
         (IDPs and primitive meteorites; middle panel) 
         and terrestrial (bottom panel) 
         carbonaceous materials.
         All exhibit ``D'' and ``G''
         (6.2, 7.7$\mum$ C--C stretching) bands
         characteristic of disordered sp$^2$-bonded {\bf aromatic} carbon.
         Taken from Sandford et al.\ (2006).
        {\bf Right (B)}: Continuum-subtracted FTIR absorption spectrum
        of a \Stardust\ sample. Both {\bf aromatic} ($\simali$3050$\cm^{-1}$)
        and aliphatic ($\simali$2967, 2928, 2872$\cm^{-1}$)
        C--H stretching bands are clearly seen.
        Taken from Keller et al.\ (2006).
        }
\end{figure}

\vspace{-7mm}
\section{~Discussion\label{sec:discussion}} 
\vspace{-3mm}
There is a long series of pieces of evidence which show that
comets have (at least partially) preserved the pristine
materials in the parent interstellar cloud out of which
the solar system has formed: 
(1) the striking similarities in the composition of 
    cometary and interstellar ices;
(2) the large deviations of the isotopic ratios
    for several elements from their terrestrial values
    (especially the high deuterium abundance);
(3) the ortho-to-para ratios of cometary water, NH$_3$,
    and CH$_4$ in several comets which imply a spin 
    temperature of $\simali$30$\K$ (which may be 
    characteristics of their formation temperature);
(4) the presence of volatiles, supervolatiles (e.g. CO),
    and rare gases (e.g. N$_2$)
    which may indicate that they were incorporated into 
    cometary nuclei at low temperatures 
    (e.g. 22$\K$ for pure CO ice);\footnote{%
      However, large-scale extensive radial mixing 
      in the solar nebula must have occurred at
      the early stage of the formation of the solar system,
      as indicated by the detection of a large number
      of crystalline olivine and pyroxene minerals
      in the \Stardust\ comet samples 
      that, based on their solar isotopic compositions,
      appear to have formed in the inner regions 
      of the solar nebula (Brownlee et al.\ 2006).
      Also, the silicate dust in the diffuse ISM is
      predominantly amorphous (Li \& Draine 2001a). 
      }
and probably also (5) the high abundance of cometary HNC
(see Crovisier 2006).
The {\it definite} detection of PAHs in comet Wild 2
by \Stardust\ provides another piece of evidence for
the connection between comets and the ISM and has
profound implications for the nature of the PAHs in
the ISM and dust disks.

Admittedly, the hypothesis of PAHs as the carrier of 
the ``UIR'' bands widely seen in the ISM (\S\ref{sec:ism})
and circumstellar dust disks (\S\ref{sec:disk})
is still a hypothesis, although the evidence in
support is very strong --
so far there is no actual precise identification 
of a single specific PAH molecule
in interstellar space or dust disks,\footnote{%
  Cernicharo et al.\ (2001) reported the detection
  of benzene (C$_6$H$_6$), the basic aromatic unit,
  in the protoplanetary nebula CRL 618.
  }
although the PAH model is quite successful in explaining 
the {\it general} pattern of band positions, relative intensities, 
and profiles observed in the ``UIR'' emission spectra,
in terms of mixtures of highly vibrationally excited neutral 
and charged PAHs. 

Details of the ``UIR'' spectra 
(precise band positions, bandwidths, and relative band intensities)
remain hard to mimic exactly 
with the use of available PAH spectra 
obtained by experimental measurements or quantum 
chemical calculations (e.g. see Fig.\,\ref{fig:pah_disks}f).
Therefore, in modeling 
the observed PAH emission spectra,
astronomers usually take an empirical approach 
by constructing ``astro-PAH'' 
absorption properties that are consistent 
with spectroscopic observations of PAH emission 
from dust 
in various astrophysical environments
(e.g. see D\'{e}sert et al.\ 1990, Schutte et al.\ 1993, 
Li \& Draine 2001, Draine \& Li 2001, 2007).
The resulting ``astro-PAH'' absorption cross sections,
although generally consistent with laboratory data 
(see Fig.\,2 of Draine \& Li 2007), 
do not represent any specific material, 
but approximate the actual absorption properties 
of the PAH mixture in astrophysical regions.

It is not surprising that the astronomical PAH 
      {\it emission} spectra
      do not closely resemble the laboratory spectrum of 
      any {\it single individual} PAH species
      since interstellar or circumstellar PAHs are most likely
      a complex mixture of many individual molecules, 
      radicals, and ions.
      As a matter of fact, Allamandola et al.\ (1999) have 
      demonstrated that the laboratory {\it absorption} spectra
      produced by co-adding different PAH spectra were 
      able to provide a detailed match to the observed 
      {\it emission} spectra.

One may still argue that the approach taken by 
      Allamandola et al.\ (1999) was not perfectly appropriate
      because they compared the astronomical {\it emission} 
      spectra with the co-added laboratory 
      {\it absorption} spectra,
      while the IR emission spectrum of a PAH
      molecule does not only depend on its IR absorption spectrum,
      but also depends on its absorption at shorter wavelengths,
      its heat capacity, and the intensity and spectral shape
      of the illuminating radiation field 
      (e.g. see Draine \& Li 2001). 
      This, together with the detection of individual
      {\it specific} PAH molecules in the \Stardust\ samples,
      suggests that it would be of great value to study
      the excitation, emission, and destruction 
      of a large number of specific PAH molecules 
      in the ISM, dust disks, and cometary comae,
      as well as the mechanism of releasing PAHs
      from the ice mantles of dust in comets
      and dust disks to the gas phase
      and their lifetime against photodestruction
      and photoionization (e.g. see Joblin et al.\ 1997,
      Li \& Lunine 2003, Li \& Draine 2007).

\vspace{3mm}
{\noindent {\bf Acknowledgements}~~ I thank J. Crovisier, 
D.E. Harker, C.M. Lisse, M.J. Mumma, and D.H. Wooden
for helpful discussions. I thank G.C. Clayton, 
S.J. Clemett, E. Furlan, V.C. Geers, M. Jura, K.H. Kim, 
S.L. Liang, B. Sargent, G.C. Sloan, and T.L. Smith
for their help in preparing for this article.
Partial support by Spitzer Theory Programs and
a HST Theory Program is gratefully acknowledged.}



\printindex

\vspace{-4mm}
\begin{thebibliography}{99.}
\vspace{-3mm}
\bibitem{}Acke, B., \& van den Ancker, M.E.\ 2004, A\&A, 426, 151 
\bibitem{}Allamandola, L.J., Tielens, A.G.G.M., \& Barker, J.R.\ 
          1985, ApJ, 290, L25
\bibitem{}Allamandola, L.J., Sandford, S.A., \& Wopenka, B.\
          1987, Science, 237, 56
\bibitem{}Allamandola, L.J., Hudgins, D.M., \& Sandford, S.A. 
          1999, ApJ, 511, L115
\bibitem{}Baas, F., Geballe, T.R., \& Walther, D.M.\ 1986, 
            ApJ, 311, L97 
\bibitem{}Beintema, D.A., et al.\ 1996, A\&A, 315, L369
\bibitem{}Beir\~ao, P., et al.\ 2006, ApJ, 643, 1
\bibitem{}Bockel\'{e}e-Morvan, D., Brooke, T.Y., \& Crovisier, J.\
            1995, Icarus, 116, 18
\bibitem{}Brooke, T.Y., Tokunaga, A.T., \& Strom, S.E.\ 
          1993, AJ, 106, 656
\bibitem{}Brownlee, D.E., et al.\ 2006, Science, 314, 1711
\bibitem{}Campins, H., \& Swindle, T.D.\ 1998, 
          Meteoritics \& Planet. Sci., 33, 1201 
\bibitem{}Cernicharo, J., et al.\ 2001, ApJ, 546, L123
\bibitem{}Clairemidi, J., Br{\'e}chignac, P., Moreels, G., 
          \& Pautet, D.\ 2004,  Planet. Space Sci., 52, 761 
\bibitem{}Clemett, S.J., Maechling, C.R., Zare, R.N., Swan, P.D., 
          \& Walker, R.M.\ 1993, Science, 262, 721
\bibitem{}Clemett, S.J., Nakamura-Messenger, K., McKay, D.S., 
          \& Sandford, S.A.\ 2007, Lunar \& Planet. Inst. Conf. Abs., 
          38, 2091 
\bibitem{}Contursi, A., et al.\ 2000, A\&A, 362, 310
\bibitem{}Coulson, I.M., \& Walther, D.M.\ 1995, MNRAS, 274, 977 
\bibitem{}Crovisier, J., 2006, Mol. Phys., 104, 2737 
\bibitem{}Crovisier, J., \& Bockel\'{e}e-Morvan, D.\
          2008, Icarus, 195, 938
\bibitem{}Crovisier, J., et al.\ 2000, 
          in ASP Conf. Ser. 196, 
          Thermal Emission Spectroscopy and Analysis of Dust, 
          Disks, and Regoliths, 109 
\bibitem{}Davies, J.K., Green, S.F., \& Geballe, T.R.\ 
          1991, MNRAS, 251, 148 
\bibitem{}Derenne, S., Rouzaud, J.N., Clinard, C., 
          \& Robert, F.\ 2005, 
          Geochimica et Cosmochimica Acta, 69, 3911 
\bibitem{}D{\'e}sert, F.X., Boulanger, F., \& Puget, J.L.\
          1990, A\&A, 237, 215 
\bibitem{}Draine, B.T., \& Li, A.\ 2001, ApJ, 551, 807
\bibitem{}Draine, B.T., \& Li, A.\ 2007, ApJ, 657, 810
\bibitem{}Draine, B.T., et al.\ 2007, ApJ, 663, 866
\bibitem{}Dwek, E.\ 2005, in AIP Conf. Proc. 761: 
          SEDs of Gas-Rich Galaxies, 103
\bibitem{}Elbaz, D., Le Floc'h, E., Dole, H., 
          \& Marcillac, D.\ 2005, A\&A, 434, L1
\bibitem{}Engelbracht, C.W., et al.\ 2005, ApJ, 628, L29 
\bibitem{}Engelbracht, C.W., et al.\ 2006, ApJ, 642, L127
\bibitem{}Flynn, G.J.\ 1989, Icarus, 77, 287 
\bibitem{}F\"{o}rster Schreiber, N. M., Roussel, H., Sauvage, M., 
          \& Charmandaris, V. 2004, A\&A, 419, 501
\bibitem{}Furlan, E., et al.\ 2006, ApJS, 165, 568 
\bibitem{}Geers, V. C., et al.\ 2006, A\&A, 459, 545 
\bibitem{}Gillett, F.C., Forrest, W.J., Merrill, K.M.\ 
          1973, ApJ, 183, 87
\bibitem{}Greenberg, J.M. 1982, in Comets, ed. L.L. Wilkening
          (Tuscon: Univ. of Arizona Press), 131 
\bibitem{}Habart, E., Natta, A., Testi, L., \& Carbillet, M.\ 
          2006, A\&A, 449, 1067
\bibitem{}Harker, D.E., Woodward, C.E., Wooden, D.H.,
          Fisher, R.S., \& Trujillo, C.\ 2007, 
          Icarus, 190, 432
\bibitem{}Higdon, S.J., Higdon, J.L., \& Marshall, J.\ 
          2006, ApJ, 640, 768 
\bibitem{}Hoban, S., Mumma, M.J., Reuter, D.C., Disanti, M., 
          Joyce, R.R., \& Storrs, A.\ 1991, Icarus, 93, 122
\bibitem{}Hudgins, D.M., \& Allamandola, L.J.\ 2005, 
          in Astrochemistry: Recent Successes 
          and Current Challenges, 443 
\bibitem{}Hunt, L., Bianchi, S., \& Maiolino, R.\ 
          2005, A\&A, 434, 849
\bibitem{}Irwin, J.A., \& Madden, S.C.\ 2006, A\&A, 445, 123
\bibitem{}Jura, M.,\ 2003, ApJ, 582, 1032
\bibitem{}Jura, M., et al.\ 2006, ApJ, 637, L45 
\bibitem{}Joblin, C., Boissel, P., \& de Parseval, P. 1997, 
          Planet. Space Sci., 45, 1539
\bibitem{}Joswiak, D.J., Brownlee, D.E., Pepin, R.O., 
          \& Schlutter, D.J.\ 2007, 
          in Dust in Planetary Systems (ESA SP-643), 141 
\bibitem{}Kaneda, H., Onaka, T., \& Sakon, I.\ 2005, ApJ, 632, L83 
\bibitem{}Keller, L.P., et al.\ 2006, Science, 314, 1728
\bibitem{}Keller, L.D., et al.\ 2008, ApJ, 684, 411
\bibitem{}L\'{e}ger, A., \& Puget, J.L.\ 1984, A\&A, 137, L5
\bibitem{}Li, A., \& Draine, B.T.\ 2001a, ApJ, 550, L213
\bibitem{}Li, A., \& Draine, B.T.\ 2001b, ApJ, 554, 778
\bibitem{}Li, A., \& Draine, B.T.\ 2008, in preparation
\bibitem{}Li, A., \& Lunine, J.I.\ 2003, ApJ, 594, 987
\bibitem{}Li, A., et al.\ 2008, in preparation
\bibitem{}Lisse, C.M., et al.\ 2006, Science, 313, 635
\bibitem{}Lisse, C.M., Kraemer, K.E., Nuth, J.A., Li, A., 
          \& Joswiak, D.\ 2007, Icarus, 187, 69
\bibitem{}Love, S.G., \& Brownlee, D.E.\ 1994, Meteoritics, 29, 69 
\bibitem{}Lutz, D., et al.\ 2005, ApJ, 625, L83
\bibitem{}Lutz, D., et al.\ 2007, ApJ, 661, L25
\bibitem{}Madden, S.C., Galliano, F., Jones, A.P., 
        \& Sauvage, M.\ 2006, A\&A, 446, 877
\bibitem{}Mann, I., Murad, E., \& Czechowski, A.\ 
          2007, Planet. Space Sci., 55, 1000 
\bibitem{}Mathis, J.S., Mezger, P.G., \& Panagia, N.\ 
          1983, A\&A, 128, 212
\bibitem{}Moutou, C., L\'{e}ger, A., D'Hendecourt, L.\
          1996, A\&A, 310, 297 
\bibitem{}Mumma, M.J., et al.\ 2001, ApJ, 546, 1183 
\bibitem{}Moreels, G., Clairemidi, J., Hermine, P., Brechignac, P., 
          \& Rousselott, P.\ 1994, A\&A, 282, 643
\bibitem{}Nier, A.O., \& Schlutter, D.J.\ 1993, 
          Meteoritics, 28, 675 
\bibitem{}O'Halloran, B., Satyapal, S., \& Dudik, R.P.\
          2006, ApJ, 641, 795
\bibitem{}Peeters, E., et al.\ 2002, A\&A, 390, 1089
\bibitem{}Peeters, E., et al.\ 2005, ApJ, 620, 774
\bibitem{}Plante, S., \& Sauvage, M.\ 2002, AJ, 124, 1995
\bibitem{}Roche, P.F., Aitken, D., Smith, C., \& Ward, M. 
          1991, MNRAS, 248, 606
\bibitem{}Rosenberg, J.L., et al.\ 2006, ApJ, 636, 742
\bibitem{}Sandford, S.A., et al.\ 2006, Science, 314, 1720
\bibitem{}Schutte, W.A., Tielens, A.G.G.M., 
          \& Allamandola, L.J.\ 1993, ApJ, 415, 397
\bibitem{}Sellgren, K.\ 2001, Spectrochimica Acta, 57, 627
\bibitem{}Sephton, M.A., et al.\ 2004, 
          Geochimica et Cosmochimica Acta, 68, 1385
\bibitem{}Siebenmorgen, R., Kr{\"u}gel, E., \& Spoon, H.W.W.\ 
          2004, A\&A, 414, 123
\bibitem{}Sloan, G.C., et al.\ 2005, ApJ, 632, 956
\bibitem{}Sloan, G.C., et al.\ 2007, ApJ, 664, 1144
\bibitem{}Smith, T.L., Clayton, G.C., \& Valencic, L.\ 
          2004, AJ, 128, 357 
\bibitem{}Smith, J.D.T., et al.\ 2004, ApJS, 154, 199 
\bibitem{}Smith, J.D.T., et al.\ 2007, ApJ, 656, 770
\bibitem{}Sylvester, R. J., et al.\ 1997, MNRAS, 289, 831
\bibitem{}Thuan, T.X., Sauvage, M., \& Madden, S.\ 
          1999, ApJ, 516, 783
\bibitem{}Tielens, A. G. G. M. 2005, The Physics and Chemistry 
          of the Interstellar Medium 
          (Cambridge: Cambridge Univ. Press) 
\bibitem{}van Boekel, R., et al.\ 2004, A\&A, 418, 177 
\bibitem{}van Diedenhoven, B., et al.\ 2004, ApJ, 611, 928
\bibitem{}van Kerckhoven, C., et al.\ 2000, A\&A, 357, 1013
\bibitem{}Verstraete, L., et al.\ 2001, A\&A, 372, 981
\bibitem{}Vijh, U.P., Witt, A.N., \& Gordon, K.D.\ 
          2005, ApJ, 633, 262
\bibitem{}Voit, G.M.\ 1991, ApJ, 379, 122 
\bibitem{}Voit, G.M.\ 1992, MNRAS, 258, 841 
\bibitem{}Werner, M.W., et al.\ 2004, ApJS, 154, 309
\bibitem{}Witt, A.N., et al.\ 2006, ApJ, 636, 303 
\bibitem{}Wooden, D.H., et al.\ 2007,
          in Protostars and Planets V, 815
\bibitem{}Wu, Y., et al.\ 2006, ApJ, 639, 157
\bibitem{}Yan, L., et al.\ 2005, ApJ, 628, 604
\bibitem{}Yan, L., et al.\ 2007, ApJ, 658, 778
\end{thebibliography}
\end{document}